\documentclass{aastex631}

\let\tablenum\relax
\usepackage{siunitx}

\shorttitle{Crab Nebula calibration of CMB polarimeters}
\shortauthors{Masi et al.}

\graphicspath{{./}{figures/}}

\begin{document}

\title{The Crab Nebula as a Calibrator for wide-beam Cosmic Microwave Background polarization surveys
}

\author[0000-0001-5105-1439]{Silvia Masi}
\affiliation{Dipartimento di Fisica, Sapienza Universit\'a di Roma and INFN sezione di Roma\\
P.le A. Moro 2, I00815 Roma, Italy}

\author[0000-0001-6547-6446]{Paolo de Bernardis}
\affiliation{Dipartimento di Fisica, Sapienza Universit\'a di Roma and INFN sezione di Roma\\
P.le A. Moro 2, I00815 Roma, Italy}

\author[0000-0002-2404-1056]{Fabio Columbro}
\affiliation{Dipartimento di Fisica, Sapienza Universit\'a di Roma and INFN sezione di Roma\\
P.le A. Moro 2, I00815 Roma, Italy}

\author{Alessandro Coppolecchia}
\affiliation{Dipartimento di Fisica, Sapienza Universit\'a di Roma and INFN sezione di Roma\\
P.le A. Moro 2, I00815 Roma, Italy}

\author{Giuseppe D'Alessandro}
\affiliation{Dipartimento di Fisica, Sapienza Universit\'a di Roma and INFN sezione di Roma\\
P.le A. Moro 2, I00815 Roma, Italy}

\author{Lorenzo Mele}
\affiliation{Dipartimento di Fisica, Sapienza Universit\'a di Roma and INFN sezione di Roma\\
P.le A. Moro 2, I00815 Roma, Italy}

\author[0000-0002-8388-3480]{Alessandro Paiella}
\affiliation{Dipartimento di Fisica, Sapienza Universit\'a di Roma and INFN sezione di Roma\\
P.le A. Moro 2, I00815 Roma, Italy}

\author[0000-0002-5444-9327]{Francesco Piacentini}
\affiliation{Dipartimento di Fisica, Sapienza Universit\'a di Roma and INFN sezione di Roma\\
P.le A. Moro 2, I00815 Roma, Italy}

\begin{abstract}
We analyze the effect of polarized diffuse emission in the calibration of wide-beam mm-wave polarimeters, when using the Crab Nebula as a reference source for both polarized brightness and polarization angle. We show that, for CMB polarization experiments aiming at detecting B-mode in a scenario with a tensor to scalar ratio $r \sim 0.001$, wide (a few degrees in diameter), precise ($\sigma_Q, \sigma_U \sim\SI{20}{\micro K_{CMB}.arcmin}$), high angular resolution ($< \mathrm{FWHM}$) reference maps are needed to properly take into account the effects of diffuse polarized emission and avoid significant bias in the calibration.
\end{abstract}


\section{Introduction} \label{sec:intro}

Current and planned searches for the linear polarization of the CMB aim at the very faint rotational component produced during cosmic inflation, the so--called $B$--mode (see e.g. \cite{Kamionkowski16}). This signature, produced by tensor perturbations originated in the very early universe,
is extremely faint. Depending on the tensor to scalar ratio $r$, currently constrained to be $r<0.06$ (see e.g. \cite{BicepKeck18, Planck20, Tristram20}), the amplitude of such a signal can be smaller than a few \si{\nano K_{CMB}}. 
Extreme {\sl sensitivity} of the polarimetric survey is thus mandatory. In addition, such a signal is embedded in a much larger (a few \si{\micro K_{CMB}}) irrotational ($E$--mode) polarized signal produced by scalar perturbations at recombination, as well as in a much larger polarized emission ($E$--mode and $B$--mode) generated by the interstellar medium in our Galaxy. This calls for extreme {\sl accuracy} of the measurements. 

Given the relative sizes of the signals, any unaccounted systematic effect (or physical process) producing $B$--mode or leaking a small fraction of the unpolarized signal or of the $E$--mode into $B$--mode would hamper this ambitious search for inflationary $B$--mode.
$B$--mode of Galactic origin can be detected and marginalized exploiting the specific spectral signatures of synchrotron emission from interstellar electrons and thermal emission from interstellar dust. 

Instrumental systematic effects must be carefully assessed and effective mitigation strategies must be put in place. Here we focus on one of the systematic effects, which is a small error $\Delta \alpha$ in the reference polarization direction of the polarimeter. This would produce a systematic bias in the $B$--mode (see e.g. \cite{Pagano09}): the presence of such a systematic error biases the measured angular auto-- and cross--power spectra $C_{\ell}'$ with respect to the original ones $C_{\ell}$ as follows:
\begin{eqnarray}\label{eq:ref_error}
C_{\ell}'^{TE} &=& C_{\ell}^{TE}\cos (2 \Delta \alpha) - C_{\ell}^{TB}\sin (2 \Delta \alpha)  ~\nonumber \\
C_{\ell}'^{TB}&=& C_{\ell}^{TE}\sin (2 \Delta \alpha) + C_{\ell}^{TB}\cos (2 \Delta \alpha) ~\nonumber \\
C_{\ell}'^{EE} &=& C_{\ell}^{EE}\cos^2 (2 \Delta \alpha) + C_{\ell}^{BB}\sin^2 (2 \Delta \alpha) - C_{\ell}^{EB}\sin (4 \Delta \alpha) ~\nonumber\\
C_{\ell}'^{BB} &=& C_{\ell}^{BB}\cos^2 (2 \Delta \alpha) + C_{\ell}^{EE}\sin^2 (2 \Delta \alpha) + C_{\ell}^{EB}\sin (4 \Delta \alpha) ~\nonumber \\
C_{\ell}'^{EB}&=& \frac{1}{2}\left(C_{\ell}^{EE}- C_{\ell}^{BB}\right)\sin (4 \Delta \alpha) + C_{\ell}^{EB}\left[\cos^2 (2 \Delta \alpha) - \sin^2 (2 \Delta \alpha)\right]
\end{eqnarray}
where $T,E,B$ stand for temperature anisotropy, $E$--mode polarization and $B$--mode polarization respectively. From these equations one can see that, in order to detect $B$--mode corresponding to $r \sim 0.001$, the error of the polarization direction reference $\Delta \alpha$ should be less than a few \si{arcmin} (see e.g. \cite{Hu03, Odea07, Miller09, Pagano09}). 

A laboratory calibration at this level of accuracy is challenging, and is not guaranteed to remain stable for the operation of the instrument in the field or during the space mission (see e.g. \cite{Pajot10}).

One possibility is to exploit the fundamental symmetries of CMB polarization, since, in the standard scenario, the expectation values for the CMB temperature to $B$--mode correlation ($\langle TB \rangle$) and $E$--mode to $B$--mode correlation ($\langle EB \rangle$) vanish. So one could use equations~(\ref{eq:ref_error}) for $C_{\ell}'^{TB}$ and $C_{\ell}'^{EB}$ to estimate $\Delta \alpha$. This approach, however, would not allow to probe non-standard scenarios, including cosmic birefringency or the effects of parity violating physics (see e.g. \cite{Carroll98, Lue99, Komatsu11, Gruppuso11, Planck16b, Molinari16}). 

A possible alternative way of measuring $\Delta \alpha$ using sky signals \citep{Minami19} exploits the fact that 
the rotation of the measured CMB polarization direction is due to both cosmic birefringency and reference direction error $\Delta\alpha$, while the rotation of galactic dust polarization is due only to $\Delta\alpha$. 

The observation of extremely well characterized linearly polarized sources in the mm-wave sky would represent the safest option for calibrating CMB polarimeters during their sky survey, in terms of the determination of the polarization angle and of the polarization degree. In this paper we analyze the problem of selecting the best polarized source in the mm-wave sky, with special attention to the problem of estimating contamination effects due to the polarized diffuse medium. The latter are especially important in forthcoming low angular resolution surveys, as LSPE \citep{Piacentini20} and LiteBIRD \citep{Hazumi20}.  

The paper is organized as follows: in section~\ref{sec:sources} we briefly summarize the properties of strong linearly polarized sources available in the mm--wave sky; in section~\ref{sec:crab} we focus on the Crab nebula and analyze the effect of diffuse foreground emission in wide beam measurements targeting the Crab, using the forthcoming LSPE
and LiteBIRD surveys as practical examples. In section~\ref{sec:conc} we discuss the results and summarize our conclusions.

\section{Linearly polarized sources in the mm--wave sky}\label{sec:sources}

A good calibration source for CMB polarization surveys should be a bright, polarization pure, stable, compact source, at high Galactic latitudes. These are rare in the mm--wave sky. The two sources producing the strongest linearly polarized signal are Taurus--A in the northern hemisphere, and Centaurus--A in the southern hemisphere. One should also mention the edge of the Moon as a strong source of linear polarized radiation at mm--waves \citep{Poppi02, QUIET11, Xu20, Yang20}. However, the power emitted is very large compared to that of the target signals of sensitive multi--frequency CMB surveys, and detector linearity issues represent a concern for the calibration. Also, in space--based surveys, the thermal stability of the instrument can be affected by the IR emission of the Moon. So we will not consider this source in the following.

The Crab Nebula (NGC 1952, Taurus--A) is a supernova remnant, with Galactic latitude $\sim -5^\circ$, emitting strong synchrotron radiation up to very high frequencies. At mm and sub--mm wavelengths, it is the brightest extra--solar source in the sky  featuring a high degree of linear polarization (with a polarized flux $\sim \SI{14}{Jy}$ at \SI{100}{GHz}) \citep{Mayer57, Allen67, Johnston69, Matveenko73, Flett79, Aumont10, Wiesemeyer11}. For this reason, Tau--A has been often considered as a polarization calibrator for precision Cosmic Microwave Background (CMB) polarization surveys \citep{Leitch02, Barkats05, Aumont10, Macias10, Weiland11, Polarbear14, Naess14, Planck16, Kusaka18, Ritacco18, Aumont20}.  

Centaurus--A (Cen--A) is an active galactic nucleus and bright radio galaxy (see e.g. \cite{Israel98}). Its Galactic latitude is 19$^\circ$. Its optical counterpart is the elliptical galaxy NGC5128, at a distance of \SI{3.8}{Mpc}. The active nucleus of the galaxy produces a rich structure of jets and plumes, well studied by radio telescopes over several orders of magnitude of angular scales.  The inner lobes of Cen--A are polarized, bright and stable at mm--waves, with a typical angular extension of \SI{10}{arcmin}. The source has been observed with CMB polarimeters (see e.g. \cite{Zemcov10}), but it is fainter than the Crab nebula (with a polarized flux $\sim \SI{2}{Jy}$ at \SI{100}{GHz}). Other even fainter sources have been considered in \cite{Burke09} for low--frequency measurements of CMB polarization. 

A different approach relies on the polarization of diffuse Galactic emission, as mapped by previous experiments (see e.g. \cite{Matsumura10}): while effective, this approach is not as direct as the measurement of a well characterized compact polarized source. The use of diffuse polarized emission, in fact, requires accurate {\sl a--priori} knowledge of the co--polar and cross--polar beams of the instrument, while in principle these can be measured directly if the polarized source is compact.

For these reasons, in the following we will focus on the Crab Nebula. 

\section{The Crab Nebula as a polarization calibrator and the effect of diffuse emission}\label{sec:crab}

We have used multi--frequency Stokes $T$, $Q$, $U$ maps from the Planck Legacy Archive\footnote{https://wiki.cosmos.esa.int/planck-legacy-archive} \citep{Dupac15, Planck20} to estimate the polarized signal measured by a wide--beam experiment targeting the Crab Nebula. The Nebula has an optical size of the order of $5^\prime\times 5^\prime$, while the beam--size of CMB experiments aiming at large--scale polarization measurements can be wider (order of $5^\prime$ to $100^\prime$ FWHM, depending on the experiment and wavelength). This means that the CMB polarimeter will detect polarized diffuse Galactic emission in addition to the polarized signal from the Crab Nebula. Since the Nebula will be diluted in the wide beam of the detector, while diffuse emission will fill it, the contribution of diffuse emission to the detected signal might be relevant. 

\begin{figure}[ht!]
\includegraphics[angle=0,scale=0.52]{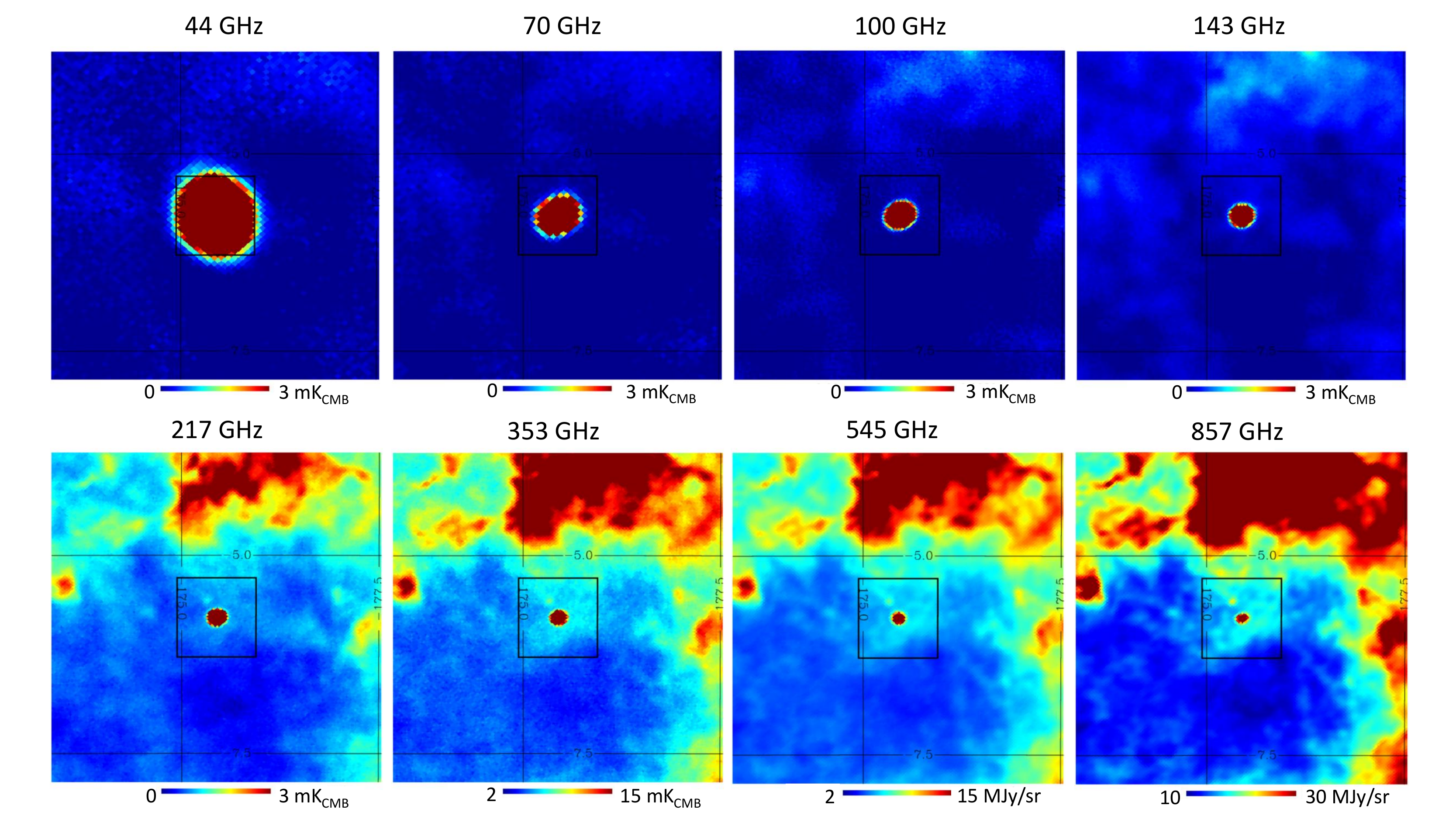}
\caption{Sky brightness measured by Planck in the regions surrounding the Crab Nebula. The maps are in Galactic coordinates, and the size of the black square centered on the Nebula is $1^\circ \times 1^\circ$. The panels are labeled with the observation frequency. The optical Nebula is not resolved at the resolution of Planck detectors. The color scale has been tuned to saturate the Nebula and display the surrounding faint diffuse emission. It is evident that its brightness increases with frequency.  \label{fig:figure1}}
\end{figure}

Inspection of the 545 and \SI{857}{GHz} intensity maps from the Planck mission confirms that the Crab Nebula lies on the same line of sight as a larger diffuse emission cloud, evident in figure~\ref{fig:figure1}. The emission is much fainter at lower frequencies, pointing to interstellar dust emission. 

In principle, one might hope to separate the emission of the nebula from diffuse interstellar emission taking advantage of their different spectral signatures. However, as a matter of facts, the Crab Nebula is evident in the foreground component separated maps from the Planck survey, both in the synchrotron map and in the dust map, and the same is true for diffuse emission. Moreover, possible rotation of the polarization angle with frequency should be properly modeled. For this reason, we argue that the spectral-based separation is not trivial, and a specialized component separation pipeline will be needed. All this is beyond the scope of this paper and will be analyzed in a future paper. Here we focus on the analysis in the different frequency bands.  

\begin{figure}[ht!]
\includegraphics[angle=180,scale=0.6]{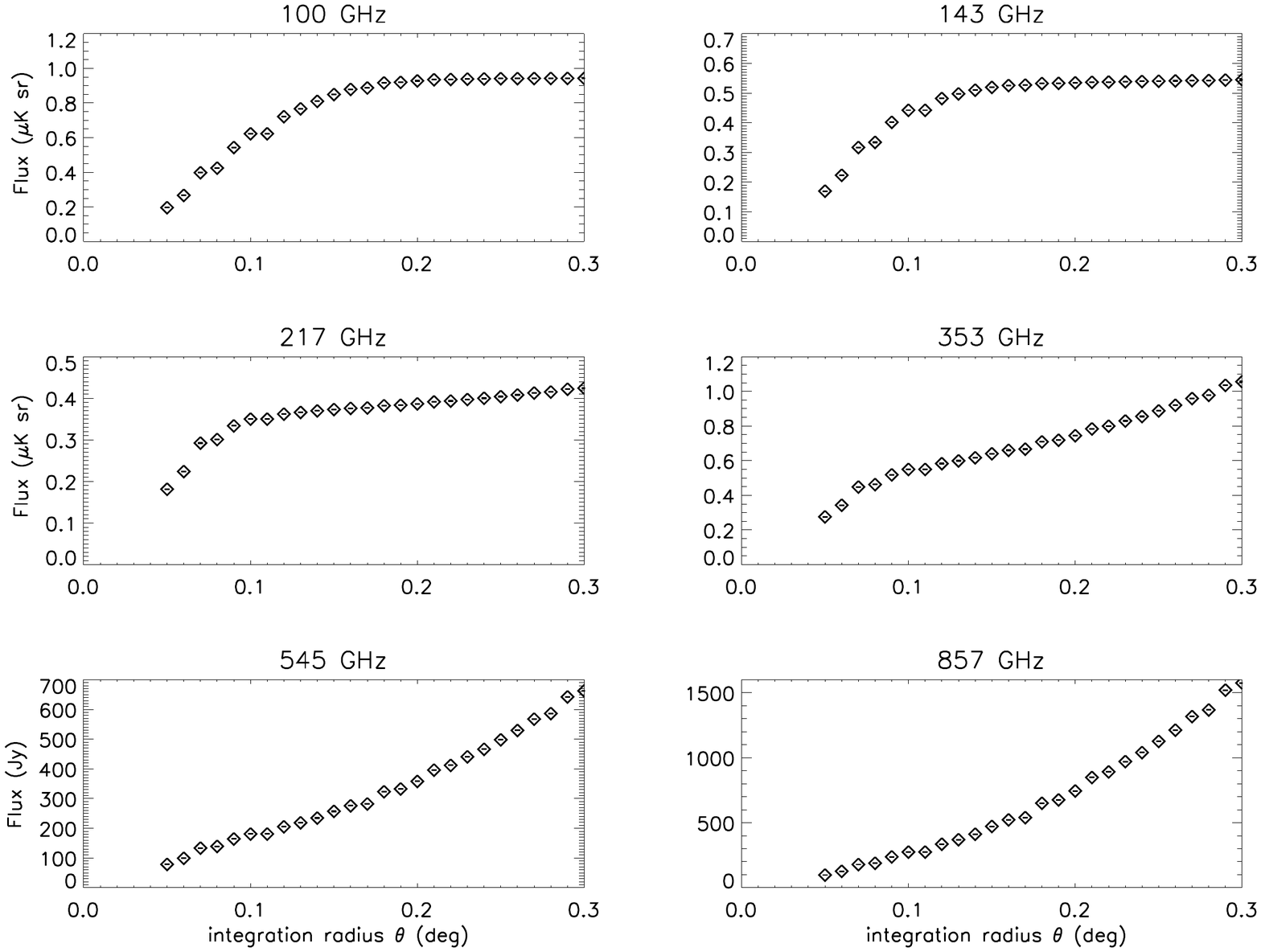}
\caption{Aperture photometry of the flux $T_m$ from the Crab Nebula (inner 5$^\prime$) and the surrounding regions, within a disk of radius $\theta$. The panels are labeled with the observation frequency.  \label{fig:figure2}}
\end{figure}

In order to estimate the relative contributions of the Crab Nebula and of diffuse polarized emission, we apply standard aperture photometry to the Planck maps for $T$, $Q$, $U$, computing the measured flux $T_m$ and the measured Stokes parameters of linear polarization $Q_m$ and $U_m$. We consider a top--hat beam with increasing radius $\theta$ and solid angle $\Omega_B = \pi \theta^2$:

\begin{eqnarray}\label{eq:fluxes}
T_m = \int_{\Omega_B} T(\ell, b) d\Omega  \\
Q_m = \int_{\Omega_B} Q(\ell, b) d\Omega  \\
U_m = \int_{\Omega_B} U(\ell, b) d\Omega 
\end{eqnarray}
Since the maps follow the Healpix pixelization scheme \citep{Gorski05} with pixel size parameter $N_{side}$, we have
\begin{equation}
    T_m = \int_{\Omega_B} T(\ell, b) d\Omega = \left[ \frac{4 \pi}{12 N_{side}^2} \right] \sum_{i=1}^n T_i 
\end{equation}
where $n$ is the number of pixels within the beam solid angle $\Omega_B$. The best estimate of the uncertainty on $F_m$ is thus
\begin{equation}\label{eq:sigmaflux} 
    \sigma^2(T_m) = \left[ \frac{4 \pi }{12 N_{side}^2} \right]^2  \sum_{i=1}^n  \sigma_i^2
\end{equation}
where $\sigma_i$ is the error of the measurement of $T$ for  pixel $i$ as obtained from the pixel covariance information in the Planck Legacy Archive files. Similar equations hold for $\sigma^2(Q_m)$ and $\sigma^2(U_m)$. We report in figure~\ref{fig:figure2} the results of the flux estimates from equation~\ref{eq:fluxes}. 

The error bars in the plots, as computed from equation~(\ref{eq:sigmaflux}), are smaller than the size of the data points. At frequencies lower than \SI{150}{GHz} the integration of the brightness converges, i.e. the flux from the Crab Nebula dominates over surrounding diffuse emission. At higher frequencies, however, the effect of diffuse emission is evident, even at distances from the center of the Crab as small as 0.15$^\circ$. 

\begin{figure}[ht!]
\includegraphics[angle=180,scale=0.6]{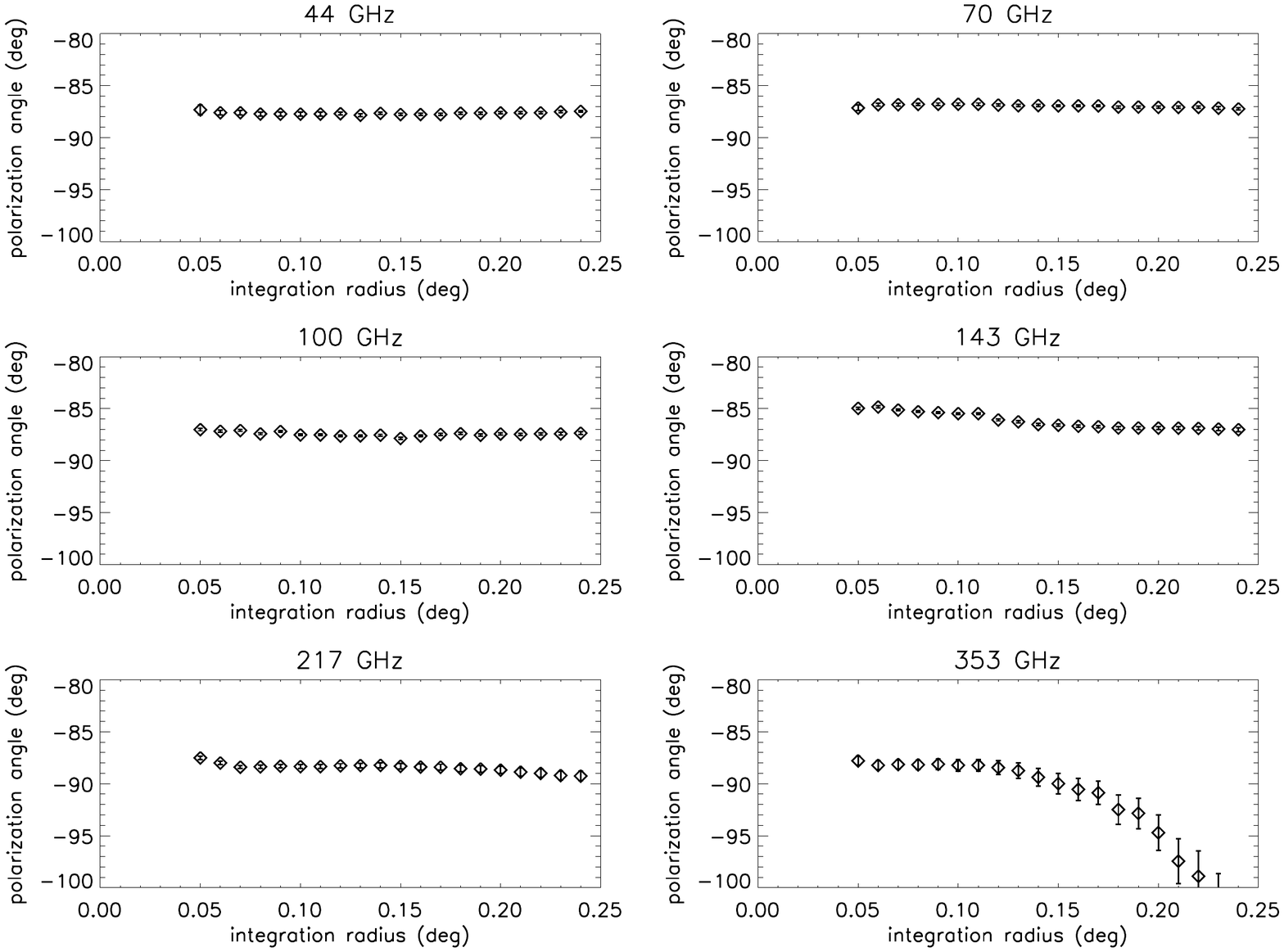}
\caption{Expected measured polarization angle $\psi_m$ integrated over the Crab Nebula area (inner 5$^\prime$) and the surrounding regions, within a disk of radius $\theta$. The panels are labeled with the observation frequency.  \label{fig:figure3}}
\end{figure}

Since interstellar emission is polarized, we expect a significant effect on the measured $Q_m$ and $U_m$ as well, especially at high frequencies. We can derive our estimates of the measured polarized signal and polarization direction, taking into account the effect of noise bias, as follows:
\begin{eqnarray}\label{eq:pipsi}
P_m = \sqrt{Q_m^2+U_m^2-\sigma^2(Q_m)-\sigma^2(U_m)} \nonumber \\
\sigma(P_m) = \frac{1}{P_m} \sqrt{Q_m^2 \sigma^2(U_m) +U_m^2 \sigma^2(Q_m)} \nonumber \\
\psi_m = \frac{1}{2} \arctan{\frac{U_m}{Q_m}}  \\
\sigma(\psi_m) = \frac{1}{2P_m^2} \sqrt{Q_m^2 \sigma^2(U_m) + U_m^2 \sigma^2(Q_m)} \nonumber
\end{eqnarray}
There is no polarization information from the Planck survey at frequencies of 545 and \SI{857}{GHz}, so we limit our estimates to lower frequencies. The most important role of the polarized source is to provide a reference polarization direction $\psi$. Published measurements of $\psi$, limited to the Crab Nebula area, are consistent with a constant angle \citep{Aumont20} $\psi_{CN} = -88.26^\circ \pm 0.27^\circ$ in Galactic coordinates. There are two issues with this result: a) this accuracy is insufficient for experiments targeting to $r \sim 0.001$, which require $\sigma(\psi) \sim$ a few \si{arcmin}; b) when the measurement beam is wider than the Crab Nebula area, additional polarized emission from the surrounding diffuse medium will be detected, and mix to the one of the Nebula area, modifying $\psi_m$ with respect to $\psi_{CN}$. In order to investigate this effect, we use Planck data and equations~(\ref{eq:pipsi}) integrating over increasing solid angles as before. The results are summarized in figure~\ref{fig:figure3}. 

For a top--hat beam wider than 20$^\prime$ FWHM, at frequencies higher than \SI{100}{GHz}, $\psi_m$ differs from $\psi_{CN}$ by a significant amount, even 1$^\circ$ or more. For frequencies lower than \SI{143}{GHz} the data do not provide strong evidence for a variation of $\psi_m$, but the errors in the estimates are still large with respect to the few $\sl arcmin$ accuracy required for detecting $r=0.001$. 

\begin{figure}[ht!]
\plotone{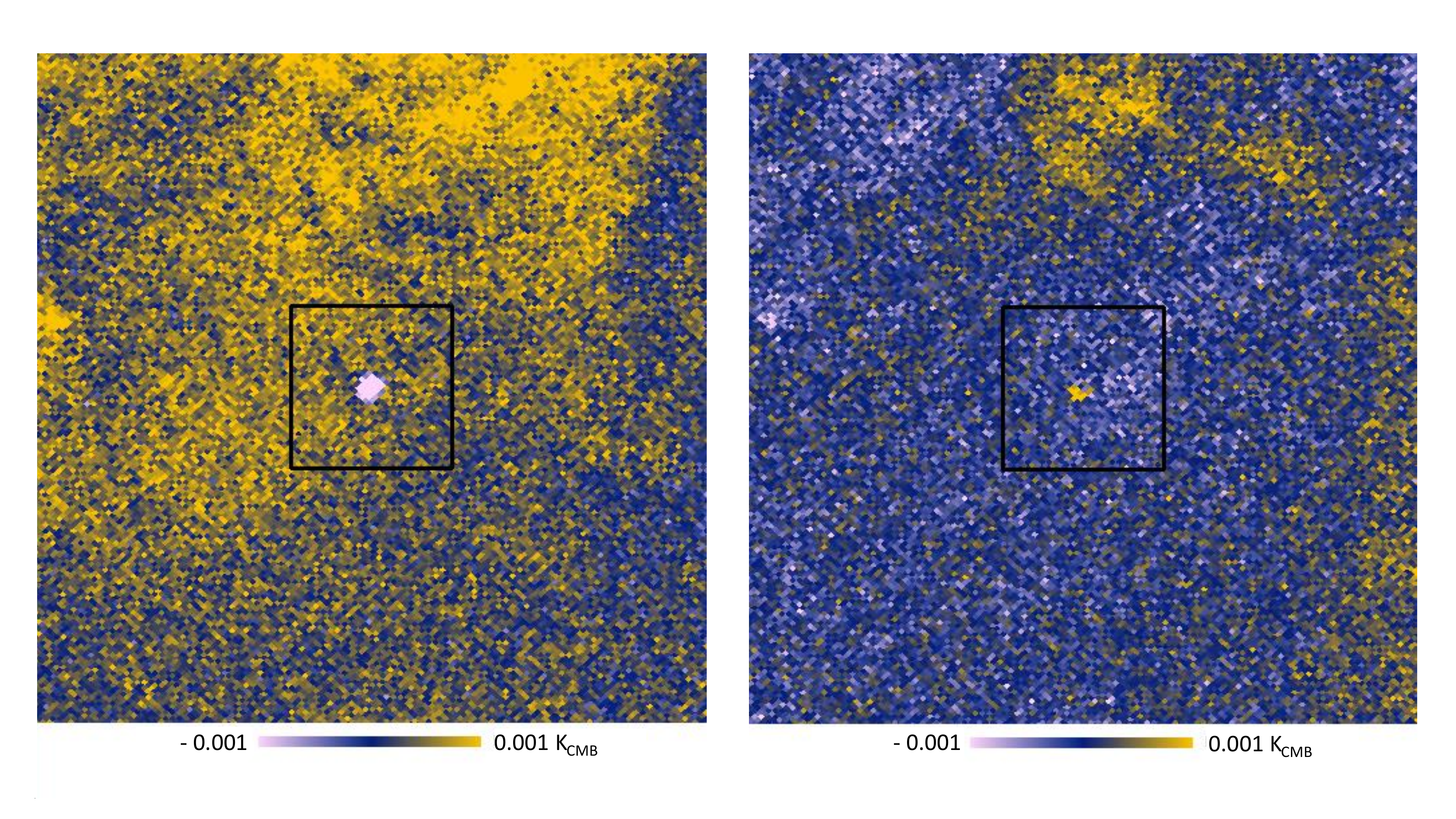}
\caption{Stokes $Q$ (left) and $U$ (right) measured by Planck at \SI{353}{GHz} in the region of the Crab Nebula. The maps are centered in the Nebula, in Galactic coordinates. The size of the black square centered on the Nebula is $1^\circ \times 1^\circ$.  \label{fig:figure4}}
\end{figure}

A concern in this kind of estimates is the effect of noise bias (see e.g. \cite{Simmons85}, \cite{Montier15}). In order to estimate the contribution of measurement noise, and to exclude bias effects in our simplistic estimators (equation~(\ref{eq:pipsi})), we carried out a simulation where the $T$, $Q$, $U$ Planck data in the Crab Nebula region have not been modified, while the Planck data in the surrounding region have been replaced by uncorrelated zero--average Gaussian noise realizations, with standard deviation mimicking the noise in the  Planck data. We repeated the calculation of $\psi_m$ on these maps, and found that the trends visible at high frequencies disappear. At \SI{353}{GHz}, the size of the fluctuations for $\psi_m$ is similar to the variations present in figure~\ref{fig:figure3} at small integration radii. Adding reasonable amounts of pixel--to--pixel correlated noise did not change the result. So we conclude that the detected deviation of $\psi_m$ from $\psi_{CN}$ for wide measurement beams is a genuine effect, due to diffuse polarized emission surrounding the Nebula. This is confirmed by visual inspection of the $Q$ and $U$ maps at \SI{353}{GHz}, reported in figure~\ref{fig:figure4}. 

From figure~\ref{fig:figure4} is evident that the Crab Nebula lays over a wide interstellar cloud with weakly positive $Q$ and slightly negative $U$. In fact, if we exclude the Nebula area from our calculations of $\psi_m$, integrating over a ring from $\theta=0.1^\circ$ to $\theta=0.25^\circ$, we obtain $\psi_{r}\left(\SI{143}{GHz}\right)\sim -95^\circ$,  $\psi_{r}\left(\SI{217}{GHz}\right)\sim -175^\circ$, $\psi_{r}\left(\SI{353}{GHz}\right)\sim -170^\circ$: these values are very far from $\psi_{CN} \sim -88^\circ$.

We have also repeated the analysis using split Planck data (first half mission versus second half mission, and even versus odd rings). We find that significant differences in $\psi_m$ are present, of the order of $\sim\SI{10}{arcmin}$ at \SI{100}{GHz} and $\sim\SI{200}{arcmin}$ at \SI{353}{GHz}.

The simple analysis carried out so far shows that if we want to use the Crab Nebula as an absolute polarization angle reference for wide--beam CMB polarization experiments:
a) we cannot use the polarization of the Crab Nebula alone, since the contribution from the surrounding regions is relevant; 
b) the accuracy of the Planck data is not sufficient to provide a reference $\psi_m$ with $\sim\si{arcmin}$ precision; 
c) we need to carry out more accurate reference measurements over a wide solid angle (at least 1$^\circ$ in diameter), with sufficient sensitivity to measure the polarized signal of the diffuse cloud in addition to the polarized signal from the Nebula itself. 
Finally, we probably need to know the details of the beam shape of the instrument under calibration, since the effect of diffuse emission will be weighted by the beam response. We focus on this in the following section. 

\section{Beam effects}

For a polarization sensitive photometer (like e.g. BICEP \cite{BICEP-IV-15}), the response to a polarized sky brightness described by the Stokes parameters $T$, $Q$, $U$ is given by:
\begin{equation}\label{eq:polsens}
V({\vec n}) = {1 \over 2}{\cal R} A \int_{4 \pi} \left[ B_T({\vec n}-{\vec n'}) T({\vec n'})  + B_Q({\vec n}-{\vec n'}) Q({\vec n'})\cos (2\psi)   + B_U({\vec n}-{\vec n'}) U({\vec n'}) \sin (2\psi)  \right]  d\Omega'
\end{equation}
where $V$ is the measured signal, ${\vec n}$ is the boresight direction, ${\cal R}$ is the responsivity of the instrument, $A$ is the collecting area; ${\vec n'}$ is the generic direction in the sky; the functions $B_T({\vec n}-{\vec n'})$, $B_Q({\vec n}-{\vec n'})$, $B_U({\vec n}-{\vec n'})$ represent the angular response (beam shape) of the instrument for unpolarized and polarized radiation; $\psi$ is the angle between the main axis of the polarization sensitive photometer and the meridian of the sky coordinates; $d\Omega'$ is the sky solid angle element. 

For a Stokes polarimeter (like e.g. SWIPE--LSPE \citep{Piacentini20} and LiteBIRD \cite{Hazumi20}), the response is given by:
\begin{equation}\label{eq:stokes}
W({\vec n}) = {1 \over 2}{\cal R} A \int_{4 \pi} \left[ B_T({\vec n}-{\vec n'}) T({\vec n'})  + B_Q({\vec n}-{\vec n'}) Q({\vec n'}) \cos (4\gamma + 2\psi)  + B_U({\vec n}-{\vec n'}) U({\vec n'}) \sin (4\gamma + 2\psi)  \right]  d\Omega'
\end{equation}
where $\gamma$ is the position angle of the half--wave plate (HWP) in the instrument reference frame. Here an ideal HWP is assumed,  which does not modify the beam pattern. 

In both cases, the beam calibration procedure is aimed at measuring the $B_T$, $B_Q$, $B_U$ responses by mapping a source with known Stokes parameters $T_s$, $Q_s$, $U_s$. If the source is compact ($\Omega_s \ll \Omega_{beam} = \int_{4 \pi} B_*({\vec n}) d\Omega$), the integrals over the solid angle in equations~(\ref{eq:polsens}) and (\ref{eq:stokes}) reduce to products:

\begin{equation}\label{eq:polsens_c}
W_c({\vec n}) = {1 \over 2} {\cal R} A \Omega_s \left[ B_T({\vec n}-{\vec n'}) T_s({\vec n'})  + B_Q({\vec n}-{\vec n'}) Q_s({\vec n'})\cos (2\psi)   + B_U({\vec n}-{\vec n'}) U_s({\vec n'}) \sin (2\psi)   \right]  
\end{equation}
\begin{equation}\label{eq:stokes_c}
W_c({\vec n}) = {1 \over 2}{\cal R} A \Omega_s \left[ B_T({\vec n}-{\vec n'}) T_s({\vec n'})  + B_Q({\vec n}-{\vec n'}) Q_s({\vec n'}) \cos (4\gamma + 2\psi)  + B_U({\vec n}-{\vec n'}) U_s({\vec n'}) \sin (4\gamma + 2\psi)  \right]  
\end{equation}

The three beam responses $B_T$, $B_Q$, $B_U$ can then be measured by separating the corresponding components of the measured signal $W_c$ by means of suitable modulation/demodulation techniques. This is the case of a ground calibration using a laboratory source, which can be rotated around the boresight to modulate $Q_s$ and $U_s$ (see e.g. \cite{Masi06, BICEP-IV-15}). In the case of a fixed sky source, as being investigated here, the required modulation is obtained by repeating the scans on the source for different angles of the polarization sensitive polarimeter around its boresight (i.e. different values of $\psi$), or repeating the scans on the source for different angles of the HWP of the Stokes polarimeter (i.e. different values of $\gamma$). 
In sky calibration of $B_T$ is obtained from observation of planets. These, however, are unpolarized sources, and do not allow to measure $B_Q$ and $B_U$. So the usual procedure consists in a ground calibration of $B_T$, $B_Q$ and $B_U$, and a sky calibration confirmation based on $B_T$ only. In the following we will assume that $B_T$, $B_Q$ and $B_U$ are known from the calibration procedure, and study how observations of a sky source with known $\psi_s$ and $P_s$ can provide a calibration of the polarization direction and of the polarimetric gain of the instrument. 

The Crab Nebula is a compact object for most of CMB polarimeters, but diffuse foreground emission, which is definitely extended, can be large enough to require the use of equations~(\ref{eq:polsens}) and (\ref{eq:stokes}), as suggested by the approximated top--hat beam analysis described in \S~\ref{sec:crab}.
Since available maps of diffuse foreground emission are noisy, the calibration will result in noisy estimates for the polarization angle reference and the polarimetric gain. Moreover, the strong emission of the Galactic plane, only $6^\circ$ away from the Crab Nebula, might contaminate significantly the detected signals, due to the intermediate sidelobes response of $B_T$, $B_Q$, $B_U$. In the following we analyze in detail two specific examples of wide--beam polarimeters, using realistic beam assumptions. 

\begin{figure}[ht!]
\includegraphics[angle=180,scale=0.6]{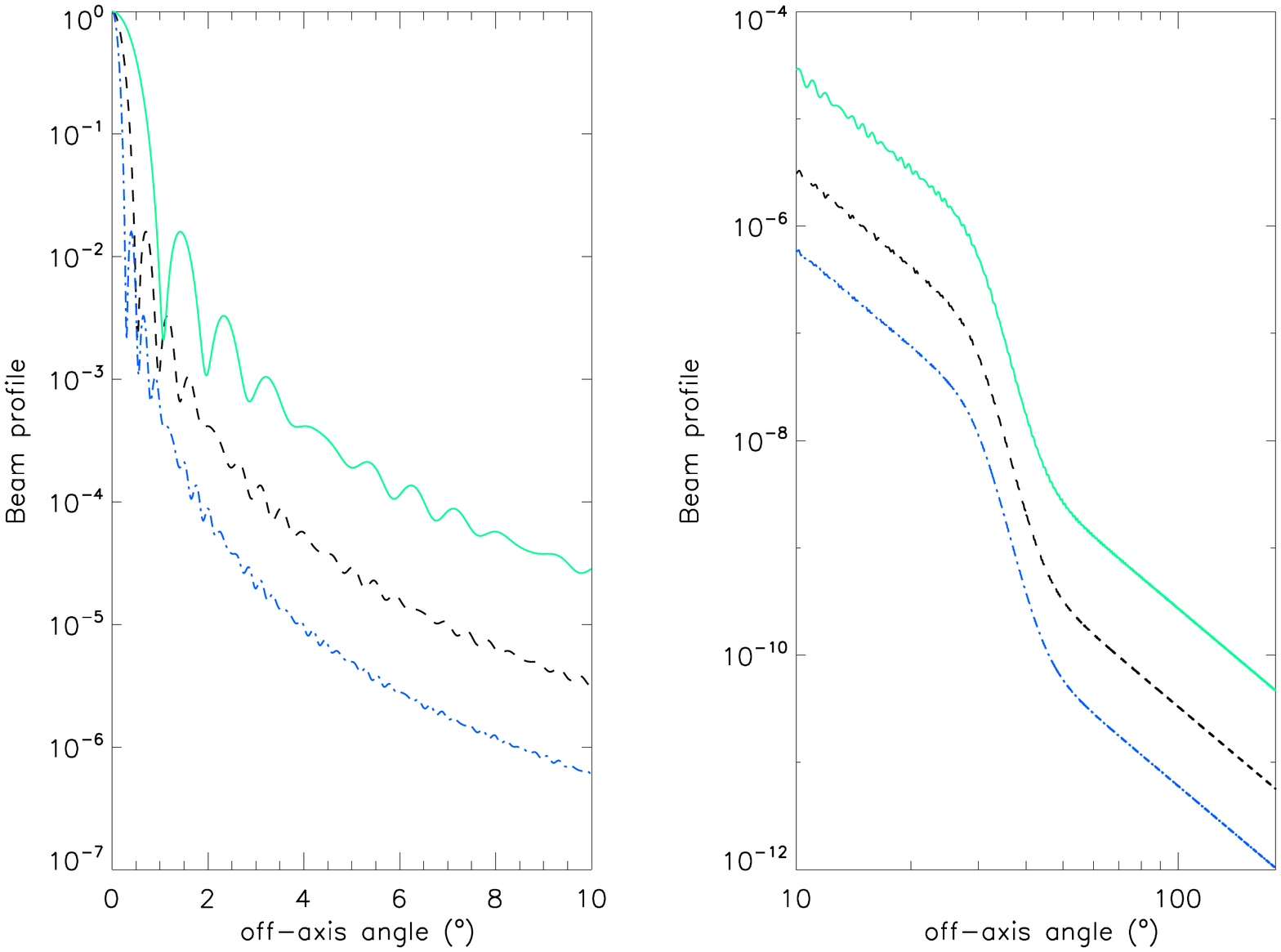}
\caption{{\bf Left:} Sample beam profiles $B(\theta)$ used in this study (equation~(\ref{eq:beam})) for single--mode instruments. The parameters ($\nu_c$, $W/\nu_c$, $D$) are as follows: (\SI{33}{GHz}, 20\%, \SI{600}{mm}) for the continuous line, (\SI{100}{GHz}, 20\%, \SI{400}{mm}) for the dashed line, (\SI{340}{GHz}, 20\%, \SI{300}{mm}) for the dot--dashed line. {\bf Right:} far sidelobes of the beams used in this study. The $\sim2$ orders of magnitude decrease of the response at $30^\circ-40^\circ$ simulates the effect of a forebaffle.
\label{fig:figure5}}
\end{figure}

For the purposes of this study we will neglect non--idealities in the polarimetric response of the instrument, and assume $B_T = B_Q  = B_U = B$. For single--mode optical systems we use a circular Airy profile averaged across the bandwidth of the instrument: 

\begin{eqnarray}\label{eq:beam}
B(\theta)={1 \over W} \int_{\nu_c-W/2}^{\nu_c+W/2} A(\theta, \nu) d\nu  \\
A(\theta, \nu) = \biggl[ {2 J_1(\rho) \over \rho} \biggr]^2  \\
\rho = \pi {D \nu \over c} \theta 
\end{eqnarray}
where $\nu_c$ is the center frequency of the receiver band, $W$ is the bandwidth, $J_1$ is the Bessel function of order 1, and $D$ is the diameter of the optical aperture of the receiver. We model the presence of a forebaffle at angle $\theta_f$ as a decrease of $B(\theta)$ for angles larger than $\theta_f$. We vary the amount of the decrease to investigate the effect of different levels of far sidelobes. Sample beams resulting from these assumptions are plotted in figure~\ref{fig:figure5}. 

These beams are assumed to be circular, while for the peripheral detectors, in a wide field--of--view array, significant ellipticity is usually measured (order of a few \%). Our circularity assumption is justified by the fact that in the following we will focus on Stokes polarimeters. Taking advantage of the polarization modulation induced by the rotation of the HWP, these polarimeters avoid the intensity to polarization leakage resulting from beam ellipticity in polarization sensitive photometers. 

In the following we will repeat the analysis carried out in \S~\ref{sec:crab}, including the effect of beam profile in the calculation of collected flux, as:
\begin{eqnarray}\label{eq:fluxes_beam}
T_{m,b} = 
\int_{0}^{2\pi} \int_{0}^{\theta_m} T(\theta, \phi) B_T(\theta) \sin \theta d\theta d\phi  \\
Q_{m,b} = 
\int_{0}^{2\pi} \int_{0}^{\theta_m} Q(\theta, \phi) B_Q(\theta) \sin \theta d\theta d\phi   \\
U_{m,b} = 
\int_{0}^{2\pi} \int_{0}^{\theta_m} U(\theta, \phi) B_U(\theta) \sin \theta d\theta d\phi 
\end{eqnarray}
Here the boresight is assumed to be in $\left[ \ell_s, b_s \right]$, i.e. the coordinates of the target source (the Crab Nebula), and the distance of the generic sky element $\left[ \ell, b \right]$ from the boresight is 
\begin{equation}
\theta = \arccos \left[ \sin b_s \sin b + \cos b_s \cos b \cos (\ell - \ell_s) \right].
\end{equation}
We study this {\sl beam-weighted} aperture photometry for the measured polarization degree $P_m$ and polarization angle $\psi_m$ as a function of the integration radius $\theta_m$, to check if the presence of diffuse polarized emission biases the results. 

\begin{figure}[ht!]
\includegraphics[angle=180,scale=0.6]{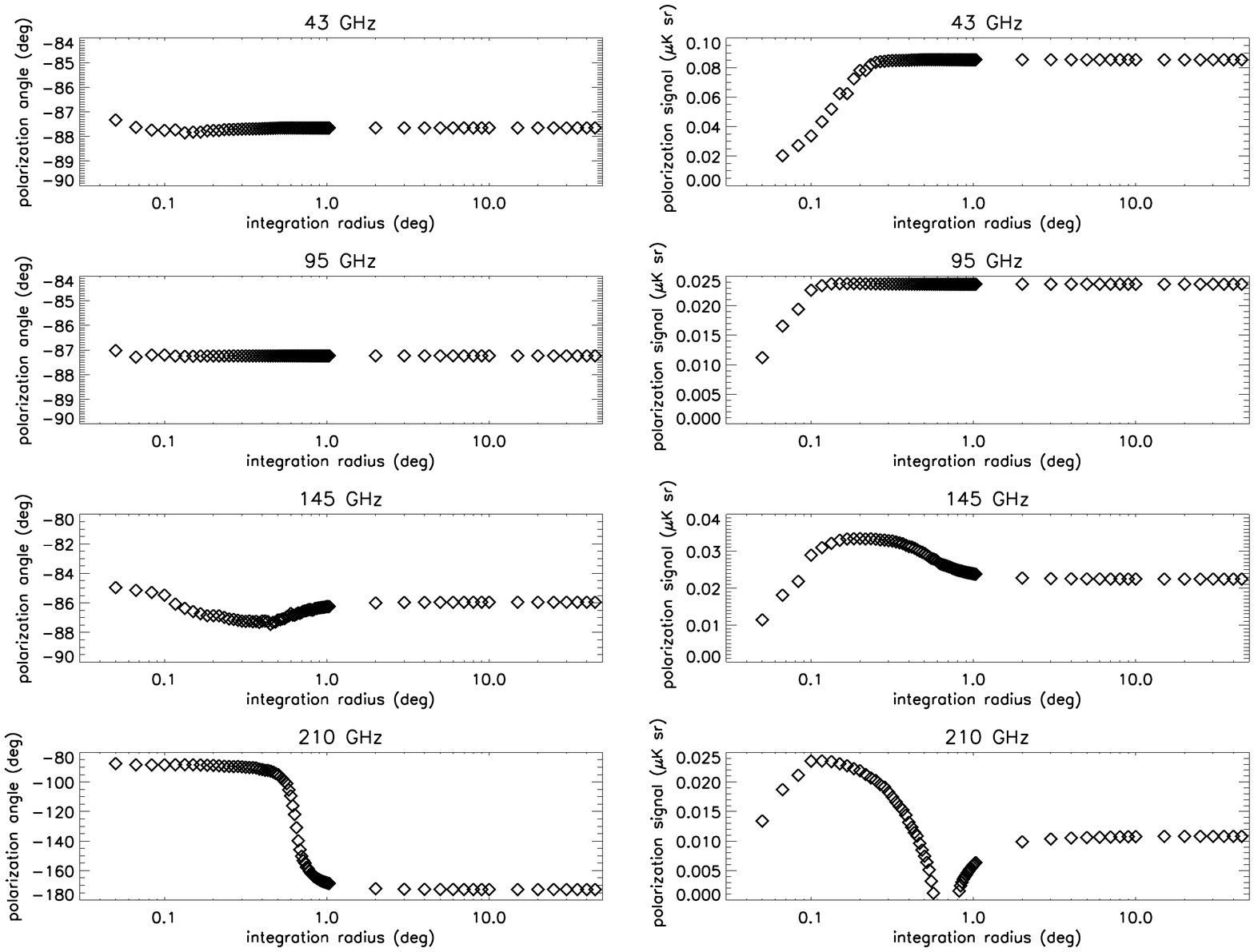}
\caption{Beam-weighted aperture photometry of the polarization angle $\psi_m$ (left) and polarized signal $P_m$ (right) for the measurements of the LSPE polarimeters at 43, 95, 145, \SI{210}{GHz} (top to bottom rows), as a function of the integration radius around the Crab Nebula. The expected measured signals correspond to the largest integration radius, and differ from the signal from the Crab Nebula alone, which corresponds to the smallest integration radius.   
\label{fig:figure6}}
\end{figure}

\subsection{LSPE}

The Large--scale Polarization Explorer (LSPE) \citep{Piacentini20} aims at the detection of B--mode in the CMB at large angular scales and includes two synergic instruments. The Short Wavelength Instrument for the Polarization Explorer (SWIPE) is a balloon--borne Stokes polarimeter, while STRIP is a ground--based coherent correlation polarimeter.

The beams of SWIPE are determined by the optical aperture of the cryogenic telescope, which is 44 cm in diameter, and by the multimode feedhorns of the detectors array \citep{Legg16}, which have the same aperture for the three measurement bands (145, 210, \SI{240}{GHz}). The beam FWHM is $\sim85^\prime$ for all bands. We use the beam responses of LSPE--SWIPE from \citep{Legg16}, modified to model also the additional sidelobes reduction from the forebaffle: the beam response is further reduced by $\sim\SI{20}{dB}$ for off--axis angles larger than $20^\circ$.

STRIP has single--mode receivers, and a large--aperture room--temperature cross--Dragone telescope, resulting in a FWHM of $20^\prime$ and $10^\prime$ for the 43 and \SI{95}{GHz} bands respectively. We used equations~(\ref{eq:beam}) as a first approximation for the beam responses of STRIP.

We compute the expected measured signals as a function of the integration radius following equation~(\ref{eq:fluxes_beam}). The results are reported in figure~\ref{fig:figure6}. We avoided to plot error bars according to equation~(\ref{eq:sigmaflux}), because the measured trends are due to intrinsic variation of the polarized sky signal, while the contribution of measurement noise is subdominant, especially for large integration radii. 

From the figure is evident that the measured angle $\psi_m$ and the polarized signal $P_m$ converge for integration radii larger than $\sim 1^\circ-2^\circ$. This means that the contribution from diffuse radiation is not negligible. As a matter of facts, $\psi_m$ and $P_m$ converge to values different from the values expected from the Crab Nebula alone. The difference is very important in the \SI{210}{GHz} and \SI{240}{GHz} channels, where the signal from the Crab Nebula dominates for integration radii smaller than $0.3^\circ$ while diffuse emission  of interstellar dust in the Galactic plane takes over at larger angles, producing a $\sim$90$^\circ$ rotation of the polarization angle. Also at lower frequencies the effect of diffuse emission is not negligible: $|\psi_m - \psi_{CN}| \sim 1^\circ$ at \SI{145}{GHz} and at \SI{43}{GHz}, while for the \SI{95}{GHz} channel $|\psi_m - \psi_{CN}| \sim 0.5^\circ$.
Larger far sidelobes rejection from the forebaffle does not modify the results significantly. 
It is thus confirmed that in order to use the Crab Nebula as an accurate reference for the calibration of SWIPE, wide reference maps (a few degrees wide, with center on the Nebula position) would be needed.

\begin{figure}[ht!]
\includegraphics[angle=180,scale=0.55]{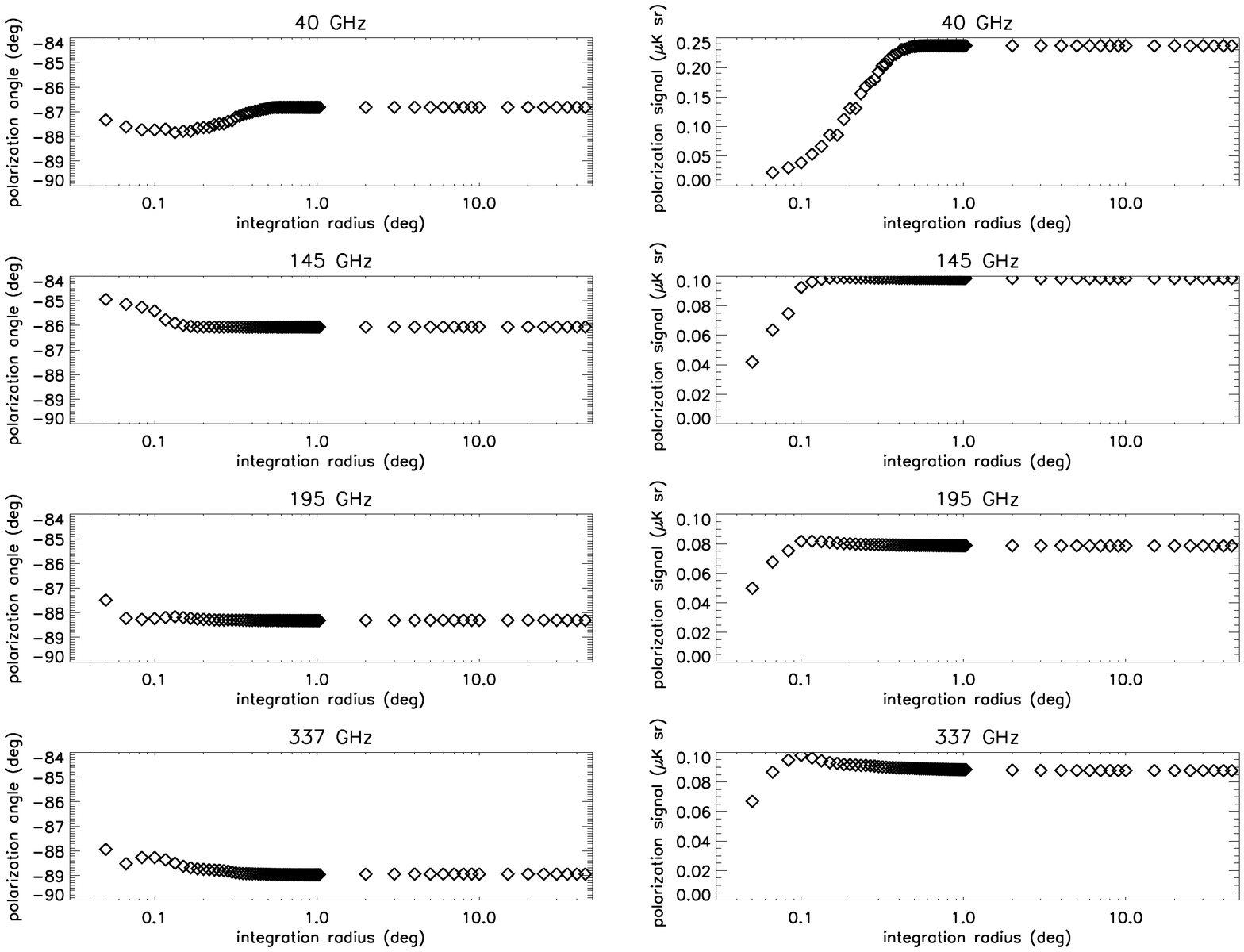}
\caption{Beam-weighted aperture photometry of the  polarization angle $\psi_m$ (left) and polarized signal $P_m$ (right) for the measurements of the LiteBIRD polarimeter at 40, 145, 195, \SI{337}{GHz} (rows from top to bottom), as a function of the integration radius around the Crab Nebula. The expected measured signals correspond to the largest integration radius, and differ from the signal from the Crab Nebula alone, which corresponds to the smallest integration radius. 
\label{fig:figure7}}
\end{figure}

\subsection{LiteBIRD}

LiteBIRD is the "Lite (Light) satellite for the studies of B--mode polarization and Inflation from cosmic background Radiation Detection", a JAXA--led space mission devoted to CMB polarization at large angular scales. LiteBIRD will produce ultrasensitive polarization maps of the full sky in 15 bands from 34 to \SI{448}{GHz} \citep{Hazumi20}. The 15 bands are distributed over 3 telescopes with different apertures. The resulting beam sizes range from $69^\prime$ to $17^\prime$ FWHM. In this case the instrument uses single--mode detectors, so we have used the beams described by equation~(\ref{eq:beam}) as a first approximation of the LiteBIRD beams.
The results for the aperture photometry described by equation~(\ref{eq:fluxes_beam}) are reported in figure~\ref{fig:figure7}.

Here the measured angle $\psi_m$ and the polarized signal $P_m$ converge for integration radii larger than $\sim 0.5^\circ$. $\psi_m$ and $P_m$ converge to values different from the values expected from the Crab Nebula alone. Due to the higher angular resolution, even at high frequencies we do not see the diffuse dust take-over seen for the LSPE-SWIPE 210 GHz channel. However, the differences are of the order of $|\psi_m - \psi_{CN}| \sim 1^\circ$. Again, it is confirmed that in order to use the Crab Nebula as an accurate reference for the calibration of LiteBIRD, wide reference maps (at least \SI{1.5}{degrees} wide, with center on the Nebula position) would be needed.

\section{Discussion and Conclusions}\label{sec:conc}

In the previous sections we have demonstrated that the effect of diffuse polarized emission can bias the result of a polarimetric calibration of a wide--field polarimeter based on observations of the Crab Nebula. Two questions now arise: 

1) how precise and accurate the reference polarization maps should be, to provide the required calibration accuracy. 

2) which is the requirement for the knowledge of the instrument beam response to attain a given calibration accuracy, which is based on convolutions of the beam responses with the polarized sky maps. 

To answer the first question we have assumed that the Planck maps (Healpix pixelization with N$_{side}=2048$) represent the real sky, in intensity and in polarization. We added 10000 realizations of uncorrelated gaussian noise to the values of T, Q, U of each pixel, with a standard deviation of the noise per pixel of \SI{10}{\micro K_{CMB}}. The resulting distribution of the $\psi_m$ angles has a standard deviation of less than \SI{1}{arcmin} for the \SI{40}{GHz} channel, and less than \SI{3}{arcmin} for the higher frequency channels. This basically sets the requirement for the depth and accuracy of the reference maps: a very demanding requirement, especially for the highest frequency maps.   

To answer the second question, we assumed again that the Planck maps represent the real sky, and  simulated a couple of typical systematic errors in the beam map $B$, investigating their impact in the measurement of $\psi_m$ and $P_m$.

The simplest systematic is an error in the solid angle of the beam. The typical measurement accuracy produces an error of the order of $<0.5\%$ on the solid angle (see e.g. \cite{PlanckBeams14}). We run our convolutions and evaluated the convergence of $\psi_m$ and $P_m$ as a function of the beam FWHM in the case of a Gaussian beam (which is a good approximation of the main beam for single--mode systems). We found that a variation of the average measured $\psi_m$ of the order of one \si{arcmin} is produced by a $\sim 1\%$ variation of the FWHM. This means that standard beam measurements are sufficient in this respect.

In principle, for this kind of measurements, errors in the evaluation of the sidelobes can be as important as errors in the estimate of the FWHM. Sidelobe response measurements become increasingly difficult as the beam response declines with the off--axis angle. For the range of off--axis angles of interest here (a few to a few tens of degrees off--axis) we expect a measurement error of  $\sim 0.1\%$ to $\sim 1 \%$ of the beam response. This is mainly due to the difficulty of controlling systematic effects, due to the large size of the device under test, including the telescope and the surrounding structures, and the implied size of the compact range illuminator and the anechoic chamber. The difficulty is exacerbated by the need of testing the full instrument at cryogenic temperature.  As an example, we have considered a 1\% to 10\% error in the amplitude of the first sidelobe of the beams used in figure~\ref{fig:figure7}, where the angular response is already at the $\sim 1 \%$ level (see left panel in figure~\ref{fig:figure5}). For a 1\% increase of the first sidelobe, the resulting change of $\psi_m$ is less than \SI{1}{arcmin}. For a 10\% increase of the first sidelobe, the change of $\psi_m$ becomes (slightly) larger than \SI{1}{arcmin} only in the \SI{337}{GHz} channel. We do not claim that this example is representative of all possible systematic effects affecting beam measurements; however it provides an indication of the required accuracy of side--lobe response determination. 

In summary, we conclude that using the Crab nebula as a reference for the calibration of future wide--beam CMB polarimeters, aiming at the measurement of B--mode with $r \sim 0.001$, is challenging, due to the effects of polarized foreground emission in the surroundings of the nebula. The measurements require wide reference maps, with a precision of $\sim \SI{20}{\micro K_{CMB}.arcmin}$, over an area with a radius of a few degrees centered on the nebula. The preparation of such set of maps requires a coordinated multi--instrument effort, including space--borne measurements at the high--frequency end of the frequency range of interest.  

\section{Acknowledgements}

This work has been supported by the Italian Space Agency (ASI) through the contracts LSPE and LiteBIRD Phase-A, and by the National Institute for Nuclear Physics (INFN) through the CSN2 activity LSPE. The paper uses data obtained with the Planck satellite (http://www.esa.int/Planck), an ESA science mission with instruments and contributions directly funded by ESA Member States, NASA, and Canada. The data are maintaned by the European Space Agency (ESA) at the the Planck Legacy Archive (https://wiki.cosmos.esa.int/planck-legacy-archive). We acknowledge the use of the HEALPix package \citep{Gorski05}. 

\bibliography{crab-pol}{}
\bibliographystyle{aasjournal}



\end{document}